\def\be{\begin{equation}}
\def\ee{\end{equation}}
\def\ba{\begin{array}}
\def\ea{\end{array}}

\documentclass[preprint,
amsmath]{revtex4}
\usepackage{float}
\usepackage[dvipsnames, svgnames, x11names]{xcolor}
\usepackage{graphicx} 
\usepackage{epstopdf}
\usepackage{lipsum}
\usepackage{caption}
\usepackage{pifont}
\usepackage{amssymb}
\def\qed{\leavevmode\unskip\penalty9999 \hbox{}\nobreak\hfill
     \quad\hbox{\leavevmode  \hbox to.77778em{%
               \hfil\vrule   \vbox to.675em%
               {\hrule width.6em\vfil\hrule}\vrule\hfil}}
     \par\vskip3pt}
\newtheorem{definition}{Definition}
\newtheorem{remark}{Remark}

\newtheorem{theorem}{Theorem}
\newtheorem{corollary}{Corollary}

\newtheorem{proposition}{Proposition}
\newtheorem{lemma}{Lemma}

\input amssym.def

\begin{document}

\title{Local Unitary Equivalence of Generic Multi-qubits Based on the CP Decomposition}
\author{Jingmei Chang$^{1, \dag}$, \ Naihuan Jing$^{2}$\\ 
$^{1}${Department of Mathematics, Shanghai University}\\  
$^{2}${Department of Mathematics, North Carolina State University}\\ 
$^\dag$ E-mail: changjingm@foxmail.com}  
\bigskip

\begin{abstract}
The CANDECOMP/PARAFAC (CP) decomposition is a generalization of the spectral decomposition of matrices to higher-order tensors. In this paper we use the CP decomposition to study unitary equivalence of higher order tensors and construct several invariants of local unitary equivalence for general higher order tensors. Based on this new method, we study the coefficient tensors of $3$-qubit states and obtain a
necessary and sufficient criterion for local unitary equivalence of general tripartite states in terms of the CP decomposition. We also generalize this method to obtain some invariants of local unitary equivalence for general multi-partite qudits.
\end{abstract}

\maketitle
\bigskip

\section{Introduction}

Local unitary (LU) equivalence is one of the key notions in quantum computation and quantum information that has wide applications such as reexportation and cryptography \cite{cy}. Extensive efforts have been devoted to finding invariants under LU transformations \cite{lu1,lu2,lu3,lu4,lu5,lu6}. Two states are completely equivalent in regards to
entanglement
if one can be transformed into the other by means of an LU transformation. In 2002
Makhlin have come up with a complete set of 18 polynomial LU invariants \cite{2Lu} in classifying  two-qubit states. After that LU invariants have also been studied for three qubit states \cite{tq}, tripartite pure \cite{r3,r4} or mixed states \cite{ms1,ms2,ms3}, arbitrary qudit states \cite{adimen} and so on.
More recently, there have been further works on LU equivalence using core tensors (see \cite{lu2}). However, the problem of multipartite LU equivalence still remains open. In this paper, we propose a new method to study the LU equivalence using the CANDECOMP/PARAFAC (CP) decomposition in
tensor analysis.

The notion of tensors, introduced in 1846 by William Rowan Hamilton, has been a
main tool to study geometry and has also been widely used in theoretical physics as far as
symmetry is concerned. Algebraically, a tensor can be understood as certain
multidimensional array, including the matrix as a special example. The higher-order equivalents of vectors (first-order) and matrices (second-order) are called higher-order tensors, multidimensional matrices, hypermatrices or multiway arrays \cite{Tencp}. Tensor decompositions were introduced by Hitchcock in 1927 \cite{td1, 1rank}, and the multiway model was studied by Cattell in 1944 \cite{Catt}. In 1981, Appellof and Davidson \cite{aD} started using
tensor decompositions in chemometrics. Recently, tensors have become even more popular
in many areas and fields, see the survey \cite{khatri} for instance. In particular, the study of tensor decomposition
has attracted a lot of attentions, and several important theories
have been developed. Among them, the CANDECOMP/PARAFAC (CP) \cite{ccp} and Tucker \cite{tucker} tensor decompositions are viewed as
 higher-order generalizations of the spectral decomposition, the singular value decomposition and the principal component analysis in matrix theory (and statistics). As the singular value decomposition and
 eigenvalues have been indispensable in quantum computation, one would also expect that
 the higher dimensional analog the CP decomposition would likely play a similar role.

 Our main goal is to introduce the CP decomposition in the study of quantum entanglement and in particular the LU equivalence problem. We will formulate some useful criteria for the LU equivalence using the CP decomposition to study multipartite quantum states. As we can see this will bring out some useful and sometimes
 operational tools to quantify density matrices in all possible scenarios. In particular, we will derive useful criteria to judge LU equivalence for tripartite and even higher-order multipartite quantum qubit states. The analysis will also suggest some tools available even for multipartite quantum qudit states, as we are dealing the
 question in complete generality.

The layout of the paper is as follows. In Sect. 2, we formulate the fundamental concept of the
CP decomposition in the context of quantum computation. In Sect. 3, some criteria of LU equivalence are given for bipartite and tripartite qubits, and some examples are provided to show how the tests are used. In Sect. 4, the criteria and tests are generalized to multi-partite quantum systems. Conclusions and summary are given in Sect. 5.

Conventions: we use bold roman letters to denote (column) vectors. For $\mathbf{a}=(a_1, \cdots, a_m)^t, \mathbf{b}=(b_1, \cdots, b_n)^t$, $\mathbf{a}\otimes\mathbf{b}=(a_1b_1, \cdots, a_1b_n, \cdots, a_mb_1, \cdots, a_mb_n)^t$. So the Kronecke product of
matrices $A=[\mathbf{a}_1, \cdots ,\mathbf{a}_m]$ and $B=[\mathbf{b}_1, \cdots ,\mathbf{b}_n]$ can be written as
$A\otimes B=[\mathbf{a}_1\otimes\mathbf{b}_1 \ \ \cdots \ \ \mathbf{a}_1\otimes\mathbf{b}_n \ \ \cdots \ \ \mathbf{a}_{m}\otimes\mathbf{b}_{1} \ \ \cdots \ \ \mathbf{a}_m\otimes\mathbf{b}_n]$.

\section{The CANDECOMP/PARAFAC decomposition of a tensor}

A tensor of format $n_1\times n_2\times \cdots \times n_N$ is a collection of numbers $x_{i_1\cdots i_N}$ from the field $\mathbf F$, where
 $1\leq i_s\leq n_s$. In general, let $V_1, \cdots, V_N$ be vector spaces over $\mathbf F$ with prescribed coordinate systems, then
a general vector in the tensor product $V_1\otimes\cdots\otimes V_N$ provides obvious configuration of order $N$ tensor.
For simplicity, we will denote $\mathcal{X}=(x_{i_{1}\cdots i_{N}})$, where $x_{i_{1}\cdots i_{N}}\in \mathbf{F}$ and
refer $\mathcal{X}\in\mathbf F^{n_1\times\cdots\times n_N}$ as a tensor. Usually, it is called an $N$-way or $N$th-order tensor, for instance, a first-order tensor is a vector $\mathbf{a}=(a_1 \ a_2 \ \cdots \ a_N)^{\mathrm{t}}$, a second-order tensor is a matrix $A=\left(\begin{smallmatrix}
a_{11} & a_{12} & \cdots & a_{1N}\\
\vdots & \vdots & \ddots & \vdots\\
a_{N1} & a_{N2} & \cdots & a_{NN}
\end{smallmatrix}\right)$, and when the order is more than two we refer them as
higher-order tensors $\mathcal{X}$. For example, a third-order tensor has three indices, and is configured as follows:

\begin{figure}[H]
\centering
\includegraphics[height=5cm,width=6cm]{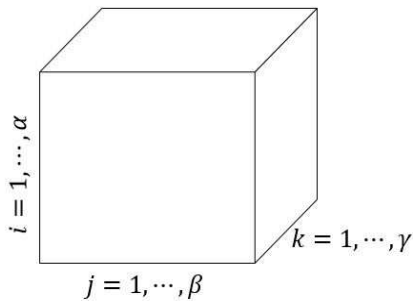}
\caption{A third-order tensor: $\mathcal{X}\in \mathbb{R}^{\alpha\times \beta\times \gamma}$.}
\label{1}
\end{figure}
There are two ways to describe tensors: fibers and slices. Fibers are higher-order analogues of matrix rows and columns, defined by fixing every index but one. A third-order tensor $\mathcal{X}$ has column, row, and tube fibers, denoted by $\mathbf{x}_{:jk}$, $\mathbf{x}_{i:k}$, and $\mathbf{x}_{ij:}$, respectively. Slices are defined by fixing all but two indices, so they are two-dimensional sections of a tensor. A third-order tensor $\mathcal{X}$ has horizontal, lateral, and frontal slices, denoted by $X_{i::}$, $X_{:j:}$, and $X_{::k}$, respectively. Simply, using $X_k$ denotes the $k$th frontal slice of a third-order tensor, $X_{::k}$. A colon is used to indicate all elements of a mode. For instance, the frontal slice of the tensor given in figure \ref{1}
is $X_k=\left(\begin{smallmatrix}
x_{11k} & x_{12k} & \cdots & x_{1\beta k}\\
\vdots & \vdots & \ddots & \vdots\\
x_{\alpha1k} & x_{\alpha2k} & \cdots & x_{\alpha\beta k}
\end{smallmatrix}\right)$, for $k=1,\cdots,\gamma$.

{\bf Matricization of a tensor:} Matricization (also called unfolding or flattening) is the process of rearranging the elements of an $N$-way array into a matrix. It is also
referred as realignment in physics literature \cite{Jing}. The mode-$n$ matricization of a tensor $\mathcal{X}\in\mathbb{R}^{\alpha_1\times \alpha_2\times\cdots\times \alpha_N}$, denoted by $X_{(n)}$, arranges the mode-$n$ fibers to be the columns of the resulting matrix.
The tensor element $x_{i_1,i_2,\dots,i_N}$ is mapped to the matrix entry $X_{(n)} (i_n,j)$ with
\begin{equation*}\label{rank1}
    j=1+\sum^{N}_{\substack{k=1 \\ k\neq n}}(i_{k}-1)\beta_k,
\end{equation*}
where $\beta_k=\prod\limits^{k-1}_{\substack{m=1 \\ m\neq n}}\alpha_{m}$.

{\bf Rank-One Tensors:} The simplest example of $N$-way tensors are outer products of $N$ vectors. Names, for $k=1, \cdots, N$, let $\mathbf{a}^{(k)}=(a^{(k)}_j)\in \mathbb R^{\alpha_k}$ be an $\alpha_k$-dimensional vector. Then
\begin{equation}\label{rank2}
    \mathcal{X}=\mathbf{a}^{(1)}\circ\mathbf{a}^{(2)}\circ\dots\circ\mathbf{a}^{(N)},
\end{equation}
 is an $N$th-order tensor in $\mathbb{R}^{\alpha_{1}\times \alpha_{2}\times\dots\times \alpha_{N}}$. Here ``$\circ$'' represents the vector outer product,
 so $x_{i_{1}i_{2}\dots i_{N}}=a^{(1)}_{i_1}a^{(2)}_{i_2}\dots a^{(N)}_{i_N}$, for all $1\leq i_{n}\leq \alpha_{n}$.
 We say an $N$th-order tensor $\mathcal X$ is
  of rank one if it can be written as an outer product of $N$ vectors as in \eqref{rank2}.

Let $\mathcal{X}\in\mathbb{R}^{I\times J}$ be a second-order tensor (i.e. a matrix) of rank $R$. Then its decomposition is
\begin{equation}\label{juzhen}
   \mathcal{X}=AB^{\mathrm{T}}=\sum^{R}_{r=1}\mathbf{a}_r\circ\mathbf{b}_r,
\end{equation}
and if the SVD of $\mathcal{X}$ is $U\Sigma V^{\mathrm{T}}$, then we can choose $A=U\Sigma$, $B=V$.

\begin{figure}[H]
\centering
\includegraphics[height=3cm,width=7cm]{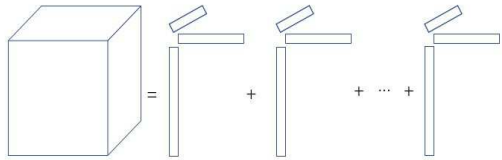}
\caption{CP decomposition of a third-order tensor.}
\label{2}
\end{figure}

For a general tensor $\mathcal X$, if $\mathcal X$ can be written as a sum of component rank-one tensors, then such a
decomposition is called a {\it CP decomposition} or canonical polyadic decomposition for $\mathcal X$. The rank $rank(\mathcal{X})$ of a tensor $\mathcal{X}$ is defined as the smallest number of rank-one tensors summing up to $\mathcal{X}$ \cite{1rank, 2rank}. In the following we usually refer a CP decomposition
with the minimal number of summands (the rank). Under a technical condition \cite{qi} such a decomposition exists but hard to compute, and there are several numerical methods to compute
(approximately) the CP decomposition.
To see a CP decomposition graphically, we use a three-way tensor to illustrate this. Suppose such a $\mathcal{X}\in \mathbb{R}^{\alpha\times \beta\times \gamma}$ has
a CP decomposition:
\begin{equation}\label{3CP}
    \mathcal{X}=\sum^{R}_{r=1}\mathbf{a}_{r}\circ\mathbf{b}_{r}\circ\mathbf{c}_{r},
\end{equation}
where $R=rank(\mathcal{X})$, $\mathbf{a}_{r}\in\mathbb{R}^{\alpha}$, $\mathbf{b}_{r}\in\mathbb{R}^{\beta}$, and $\mathbf{c}_{r}\in\mathbb{R}^{\gamma}$ for $r=1,\dots,R$. Then the elements of $\mathcal{X}$ in Eq.(\ref{3CP}) are written as
\begin{equation}\label{3CE}
    x_{ijk}=\sum^{R}_{r=1}a_{ir}b_{jr}c_{kr},
\end{equation}
which is illustrated in Fig. \ref{2}.

For $\mathcal X$ given in \eqref{3CP}, 
the {\it factor matrices}
are defined by $A=[\mathbf{a}_{1} \ \mathbf{a}_{2} \ \cdots \ \mathbf{a}_{R}]$, $B=[\mathbf{b}_{1} \ \mathbf{b}_{2} \ \cdots \ \mathbf{b}_{R}]$,
$C=[\mathbf{c}_{1} \ \mathbf{c}_{2} \ \cdots \ \mathbf{c}_{R}]$. Then
tensor $\mathcal{X}$ can be represented in the following matricized form:
\begin{equation}\label{matricization}
    X_{(1)}=A(C\odot B)^\mathrm{t}, \ \
    X_{(2)}=B(C\odot A)^\mathrm{t}, \ \
    X_{(3)}=C(B\odot A)^\mathrm{t},
\end{equation}
where $\odot$ stands for the Khatri-Rao product, which is defined for $P\in\mathbb R^{\alpha\times \gamma}, Q\in\mathbb R^{\beta\times \gamma}$
$P\odot Q=[\mathbf{p}_1\otimes\mathbf{q}_1 \ \mathbf{p}_2\otimes\mathbf{q}_2 \ \cdots \ \mathbf{p}_{\gamma}\otimes\mathbf{q}_{\gamma}]\in\mathbb R^{\alpha\beta\times \gamma}$.
In general, the horizontal and lateral slices have the similar equations. In such a situation, we write the CP decomposition concisely as follows (following Kolda):
\begin{equation}\label{3CPm}
    \mathcal{X}=\sum^{R}_{r=1}\mathbf{a}_{r}\circ\mathbf{b}_{r}\circ\mathbf{c}_{r}=[\![A,B,C]\!],
\end{equation}
where $A, B, C$ are the factor matrices.

{For convenience, one can normalize the columns of $A$, $B$ and $C$ as length one. Then, we can write that 
\begin{equation}\label{norCP}
\mathcal{X}=\sum^{R}_{r=1}\lambda_r\mathbf{\hat{a}}_{r}\circ\mathbf{\hat{b}}_{r}\circ\mathbf{\hat{c}}_{r}=[\![\mathbf{\lambda};\hat{A},\hat{B},\hat{C}]\!],
\end{equation}
where $R$ is the rank of tensor $\mathcal{X}$ and $\mathbf{\lambda}\in\mathbb{R}^{R}$. The mode-$n$ matricized versions are then $X_{(1)}=\hat{A}\Lambda(\hat{C}\odot \hat{B})^{\mathrm{t}},\ \ X_{(2)}=\hat{B}\Lambda(\hat{C}\odot \hat{A})^\mathrm{t}, \ \ X_{(3)}=\hat{C}\Lambda(\hat{B}\odot \hat{A})^\mathrm{t},$ with $\Lambda=diag(\mathbf{\lambda})$. }

As we remarked earlier that the tensor $\mathcal{X}$ always admits such a tensor CP decomposition \cite{qi}.
Now let's discuss the uniqueness of the CP decomposition \cite{Tencp,unicpd}. 
If a tensor $\mathcal X$ of rank $R$ has a CP decomposition with $R$ summands, then it is essentially unique in the following sense.
Suppose $\mathcal X$ has two decompositions with factor
 matrices $(A,B,C)$ and $(\bar{A},\bar{B},\bar{C})$, then under a mild
condition 
\begin{eqnarray*}\label{eq}
    \bar{A}=A\Pi\Delta_1, \
    \bar{B}=B\Pi\Delta_2, \
    \bar{C}=C\Pi\Delta_3,
\end{eqnarray*}
where $\Pi$ is a $R\times R$ permutation matrix, which corresponds to a reordering of the rank-one component tensors,
and ${\Delta}_{i}=\mathrm{diag}(d^i_{11}\cdots d^i_{RR})$, $i=1,2,3$, are diagonal matrices with ${\Delta}_1{\Delta}_2{\Delta}_3={\mathrm{I}}_R$, i.e. $d^1_{rr}d^2_{rr}d^3_{rr}=1$, $r=1,\cdots,R$.

The permutation indeterminacy refers to the equation
\begin{equation}\label{pin}
    \mathcal{X}=[\![{A},{B},{C}]\!]=[\![{A\Pi},{B\Pi},{C\Pi}]\!],
\end{equation}
and means that the rank-one component tensors can be reordered arbitrarily.
The scaling indeterminacy refers to the equation
\begin{equation}\label{sin}
    \mathcal{X}=\sum^{R}_{r=1}(d^1_{rr}\mathbf{a}_{r})\circ(d^2_{rr}\mathbf{b}_{r})\circ(d^3_{rr}\mathbf{c}_{r}),
\end{equation}
where the individual vectors can be scaled as long as the scaling factors are multiplied to $1$ .

Let $A\in \mathbb{C}^{I\times J}$, we shall use the following definitions \cite{2rank,JR1}:

(i) $r_A=rank(A)=r$ if and only if it contains at least a collection of $r$ linearly independent columns, and this fails for $r+1$ columns.

(ii) $k_A=r$ which denotes the $k$-rank of $A$, if and only if every $r$ columns are linearly independent, and this fails for every $r+1$ columns.

Therefore $k_A\leq r_A\leq\min(I,J),\ \forall A$.

Based on equation (\ref{3CPm}), Kruskal \cite{JR1,2rank} gave a sufficient condition for essential uniqueness of the CP decomposition with factor matrices ${A}$, ${B}$, ${C}$:
\begin{equation}\label{sc}
k_A+k_B+k_C\geq 2R+2.
\end{equation}

We recall that
the Hadamard product for matrices $A,B\in\mathbb{R}^{\alpha\times\beta}$ is defined by 
$A\ast B=(a_{ij}b_{ij})\in\mathbb R^{\alpha\times\beta}$.
Together with the Kronecker and Khatri prodcts, they satisfy the following properties \cite{proper,khatri}:
$(A\otimes B)(C\otimes D)=AC\otimes BD$, $(A\otimes B)^{\dag}=A^{\dag}\otimes B^{\dag}$, $(A\odot B)^{t}(A\odot B)=A^{t}A\ast B^{t}B$.


If $\mathcal X=(x_{i_1\dots i_N})$ is a tensor of format $n_1\times\cdots\times n_N$. We can define the action of $\mathrm{GL}_{n_1}\otimes\cdots
\otimes \mathrm{GL}_{n_N}$ as follows. For $A_k=(a_{i_kj_k})\in\mathrm{GL}_{n_k}$, the product $(A_1\otimes \cdots \otimes A_N)\mathcal X$
is defined by
\begin{equation}
\left((A_1\otimes \cdots \otimes A_N)\mathcal X\right)_{i_1i_2\dots i_N}=\sum\limits_{k_1,k_2,\dots, k_N} a_{i_1k_1}a_{i_2k_2}\cdots a_{i_Nk_N}x_{k_1k_2\dots k_N}
\end{equation}

\section{LU invariants of generic 3-qubit states under tensor's CP decomposition}

Let $\rho$ be a $3$-qubit state on a $2\times 2\times 2$ dimensional Hilbert space $H(=H^2_1\otimes H^2_2\otimes H^2_3)$:
\begin{equation}\label{rho}
\rho=\sum^3_{i_1,i_2,i_3=0}x_{i_1i_2i_3}\sigma^{(1)}_{i_1}\otimes\sigma^{(2)}_{i_2}\otimes\sigma^{(3)}_{i_3},
\end{equation}

\begin{equation}\label{x}
x_{i_1i_2i_3}={\frac{1}{2^3}}\mathrm{Tr}(\rho\sigma^{(1)}_{i_1}\otimes\sigma^{(2)}_{i_2}\otimes\sigma^{(3)}_{i_3})
\end{equation}
are real coefficients.

\begin{definition}\label{xii}
Let $\mathcal{X}_{j_1\cdots j_M}(M\leq 3,1\leq j_1<\cdots<j_M\leq 3)$ be the $M$th-order tensor composed of elements $x_{i_1i_2i_3}$ where the subscript is zero except for the $i_{j_1}$-th, $\cdots$, $i_{j_M}$-th position.
\end{definition}

For convenience, we call them coefficient tensors of the state when considering the tensor such as \eqref{x}.
For example, $\mathcal{X}_2=(x_{0 i_{2}0})=(x_{010},x_{020},x_{030})$ is a three dimensional vector, $\mathcal{X}_{13}=(x_{i_{1}0i_{3}})=\begin{pmatrix}
     x_{101} & x_{102} & x_{103} \\
     x_{201} & x_{202} & x_{203} \\
     x_{301} & x_{302} & x_{303}
\end{pmatrix}$ is a $3\times 3$ matrix, and $\mathcal{X}_{123}=(x_{i_1i_2i_3})$ is a third-order tensor, where $i_1,i_2,i_3=1,2,3$.

Two $N$-qubit mixed states $\rho$ and $\rho'$ are {\it local unitary equivalent} if
\begin{equation}\label{lu}
\rho'=(U_1\otimes\cdots\otimes U_N)\rho(U_1\otimes\cdots\otimes U_N)^{\dag}
\end{equation}
for some $U_i\in SU(2),i=1,2,\cdots,N$. Here $\dag$ denotes transpose and conjugate.

The following is well-known \cite{JYZ}.
\begin{lemma}\label{l} Let $U\in SU(2)$, $\sigma_{k}(k=1,2,3)$ are the Pauli matrices. 
Then \be\label{sou}
U\sigma_{i}U^{\dag}=\sum^{3}_{j=1}O_{ji}\sigma_{j},
\ee
\end{lemma}
where $i=1,2,3$, and $(O_{ij})\in SO(3)$.


%


\begin{lemma}\label{e} Two generic 3-qubit states $\rho$ and $\rho'$ are local unitary equivalent if and only if there are real orthogonal matrices $O_1, O_2, O_3\in SO(3)$ such that
\be\label{lemma 2}
(O_{j_1}\otimes\cdots\otimes O_{j_M})\mathcal{X}_{j_1\cdots j_M}=\mathcal{X}'_{j_1\cdots j_M},
\ee
where $M\leq 3,1\leq j_1<\cdots<j_M\leq 3$, $\mathcal{X}_{j_1\cdots j_M}$ and $\mathcal{X}'_{j_1\cdots j_M}$ are the coefficient tensors
of $\rho$ and $\rho'$ respectively.
\end{lemma}

\noindent
{\bf{Proof:}} Let $\rho=\sum\limits^{3}_{i_1,i_2,i_3=0}x_{i_1i_2i_3}\sigma^{(1)}_{i_1}\otimes\sigma^{(2)}_{i_2}\otimes\sigma^{(3)}_{i_3}$.
Note that $U_i[\sigma_1, \sigma_2, \sigma_3]U_i^{\dagger}=[\sigma_1, \sigma_2, \sigma_3]O_i^t$, where $O_i\in O(3)$ by Lemma \ref{1}. Then
$\rho'=(U_1\otimes U_2\otimes U_3)\rho(U_1\otimes U_2\otimes U_3)^{\dag}$ implies that
\begin{equation}\label{lxx'}
x'_{klm}=\sum^{3}_{i_1,i_2,i_3=0}x_{i_1i_2i_3}O^{(1)}_{ki_1}O^{(2)}_{li_2}O^{(3)}_{mi_3},
\end{equation}
where
\begin{align*}
\begin{split}
O^{(j)}_{si_j}= \left \{
\begin{array}{ll}
1, &s=0,i_j=0 \\
0, &s=0, i_j\neq0 \ or \ s\neq0, i_j=0,
\end{array}
\right.
\end{split}
\end{align*}
$s=0,1,2,3$ and $j=1,2,3$.
Therefore $(O_{j_1}\otimes\cdots\otimes O_{j_M})\mathcal{X}_{j_1\cdots j_M}=\mathcal{X}'_{j_1\cdots j_M}$. The converse follows from the fact that $\mathrm{SU}(2)$ is a double cover of $\mathrm{SO}(3)$.
$\hfill\Box$




Now we show that the rank of the tensors for two states remain the same under local unitary transformation.

\begin{lemma}\label{mm} Let $\mathcal{X}_{j_1\cdots j_M}$ and $\mathcal{X}'_{j_1\cdots j_M}$ be the coefficient
 tensors of local unitary equivalent density matrices
$\rho$ and $\rho'$ respectively, then
Rank$(\mathcal{X}_{j_1\cdots j_M})$=Rank$(\mathcal{X}'_{j_1\cdots j_M})$.
\end{lemma}
{\bf{Proof:}} If $R_{j_1\cdots j_M}$ is the rank of the coefficient
tensor $\mathcal{X}_{j_1\cdots j_M}$ of $\rho$, then $\mathcal{X}_{j_1\cdots j_M}$ can be written as
\be\label{lemma 3X}
\mathcal{X}_{j_1\cdots j_M}=\sum^{R_{j_1\cdots j_M}}_{r=1}\mathbf{a}^{({j_1}\setminus j_1\cdots j_M)}_{r}\circ\cdots\circ\mathbf{a}^{({j_M}\setminus j_1\cdots j_M)}_{r},
\ee
where ${j_k}\setminus j_1\cdots j_M$ represents the part $k$ of the CP decomposition about tensor $\mathcal{X}_{j_1\cdots j_M}$, $\mathbf{a}^{({j_k}\setminus j_1\cdots j_M)}_{r}\in\mathbb{R}^3$, $M\leq 3,1\leq j_1<\cdots<j_M\leq 3,k=1,\cdots,M$.
Since $\rho'$ is local unitary equivalent to $\rho$, there are orthogonal matrices $O_{j_M}\in \mathrm{SO}(3)$ such that
\be\label{lemma 3X'}
\mathcal{X}'_{j_1\cdots j_M}=(O_{j_1}\otimes\cdots\otimes O_{j_M})\mathcal{X}_{j_1\cdots j_M}=\sum^{R_M}_{r=1}(O_{j_1}\mathbf{a}^{({j_1}\setminus j_1\cdots j_M)}_{r})\circ \cdots\circ (O_{j_M}\mathbf{a}^{({j_M}\setminus j_1\cdots j_M)}_{r})
\ee
Therefore $R'_{j_1\cdots j_M}\leq R_{j_1\cdots j_M}$. By exchanging $\rho$ and $\rho'$, we also have $R_{j_1\cdots j_M}\leq R'_{j_1\cdots j_M}$. Hence $R'_{j_1\cdots j_M}=R_{j_1\cdots j_M}$.
$\hfill\Box$

\begin{theorem}\label{t} Two generic 3-qubit states $\rho$ and $\rho'$ are local unitary equivalent if and only if
there exist the corresponding factor matrices of $\mathcal{X}_{j_1\cdots j_M}$ and $\mathcal{X}'_{j_1\cdots j_M}$ in
the same orbit under the action of $\mathrm{SO}(3)^{\otimes 3}$, i.e.
if $[\![A^{j_1\cdots j_M}_{j_1}, \cdots, A^{j_1\cdots j_M}_{j_M}]\!]$ is the factor matrix of $\mathcal{X}_{j_1\cdots j_M}$, then
$\rho$ and $\rho'$ are local unitary equivalent if and only if there is a factor matrix $[\![ A'^{j_1\cdots j_M}_{j_1}, \cdots, A'^{j_1\cdots j_M}_{j_M}]\!]$ of $\mathcal{X}'_{j_1\cdots j_M}$ such that
\begin{equation}\label{leq}
{A}'^{j_1\cdots j_M}_{j_1}=O_{j_1}{A}^{j_1\cdots j_M}_{j_1},  \quad  \cdots, \quad {A}'^{j_1\cdots j_M}_{j_M}=O_{j_M}{A}^{j_1\cdots j_M}_{j_M},
\end{equation}
where $M\leq 3$, $1\leq j_1<\cdots<j_M\leq 3$, and $O_{j_k}\in \mathrm{SO}(3),k=1,\cdots,M$.
\end{theorem}

\noindent
{\bf{Proof:}} 
Suppose each $\mathcal X_{j_1\cdots j_M}$ for $\rho$ has the CP decomposition
\begin{equation}\label{lx}
\mathcal{X}_{j_1\cdots j_M}=\sum^{R_{j_1\cdots j_M}}_{r=1}\mathbf{a}^{({j_1}\setminus j_1\cdots j_M)}_{r}\circ\cdots\circ\mathbf{a}^{({j_M}\setminus j_1\cdots j_M)}_{r}
=[\![A^{j_1\cdots j_M}_{j_1}, \cdots, A^{j_1\cdots j_M}_{j_M}]\!],
\end{equation}
where the $R_{j_1\cdots j_M}=rank(\mathcal{X}_{j_1\cdots j_M})$.

It follows from Lemma \ref{mm} and Lemma \ref{e} that there exist $O_i\in\mathrm{SO}(3)$ such that $(O_{j_1}\otimes\cdots\otimes O_{j_M})\mathcal{X}_{j_1\cdots j_M}=\mathcal{X}'_{j_1\cdots j_M}$, i.e.
\begin{equation}\label{th1}
\mathcal{X}'_{j_1\cdots j_M}=(O_{j_1}\otimes\cdots\otimes O_{j_M})\mathcal{X}_{j_1\cdots j_M}=\sum^{R_{j_1\cdots j_M}}_{r=1}(O_{j_1}\mathbf{a}^{({j_1}\setminus j_1\cdots j_M)}_{r})\circ \cdots\circ (O_{j_M}\mathbf{a}^{({j_M}\setminus j_1\cdots j_M)}_{r}).
\end{equation}
Therefore, $[\![O_{j_1}A^{j_1\cdots j_M}_1, \cdots, O_{j_M}A^{j_1\cdots j_M}_M]\!]=[\![A'^{j_1\cdots j_M}_1, \cdots, A'^{j_1\cdots j_M}_M]\!]$ is the factor matrix of $\rho'$. The reverse direction can be shown similarly. $\hfill\Box$

\begin{theorem}\label{h} Let $\rho$ be a generic 3-qubit state with the coefficient tensor $\mathcal{X}$ having the factor matrices
$A^{j_1\cdots j_M}_{j_1}, \cdots, A^{j_1\cdots j_M}_{j_M}$.
A 3-qubit state $\rho'$ is local unitary equivalent to $\rho$
if and only if there exist $A'^{j_1\cdots j_M}_{j_s}(s=1,2,\cdots,M)$, the corresponding $\mathcal{X}'_{j_1\cdots j_M}$ of $\rho'$ has the CP decomposition:
\be\label{T2X}
\mathcal{X'}_{j_1\cdots j_M}=\sum^{R_{j_1\cdots j_M}}_{r=1}\mathbf{a'}^{(j_1)}_{r}\circ\cdots\circ\mathbf{a'}^{(j_M)}_{r}=[\![{A'^{j_1\cdots j_M}_{j_1}},\cdots,{A'^{j_1\cdots j_M}_{j_M}}]\!]
\ee
such that
\begin{eqnarray}\label{T2X'1}
&&(A^{j_1\cdots j_M}_{j_i})^{\mathrm{t}}A^{j_1\cdots j_M}_{j_i}=(A'^{j_1\cdots j_M}_{j_i})^{\mathrm{t}}{A}'^{j_1\cdots j_M}_{j_i}, \\
\label{T2X'2}&&(A^{123}_{j_i})^{\mathrm{t}}{A^{j_1\cdots j_M}_{j_i}}=(A'^{123}_{j_i})^{\mathrm{t}}{A}'^{j_1\cdots j_M}_{j_i},
\end{eqnarray}
where $\mathrm{t}$ denotes transpose, $M\leq 3,1\leq j_1<\cdots<j_M\leq 3$, and $i=1,\cdots,M$.
\end{theorem}

\noindent
{\bf{Proof:}} One direction is clear from Theorem \ref{t}. We are left to
verify that Eq.(\ref{T2X'1}) and Eq.(\ref{T2X'2}) will imply that $\rho$ and $\rho'$
are local unitary equivalent.

Suppose \eqref{T2X'1} holds. Take $(A^{j_1\cdots j_M}_{j_i})^{\mathrm{t}}A^{j_1\cdots j_M}_{j_i}=(A'^{j_1\cdots j_M}_{j_i})^{\mathrm{t}}{A}'^{j_1\cdots j_M}_{j_i}$,
by the thin version of the singular value decomposition and \cite[Theorem 7.3.11]{th},
there exists $O^{j_1\cdots j_M}_{j_i}\in \mathrm{SO}(3)$ such that ${A}'^{j_1\cdots j_M}_{j_i}=O^{j_1\cdots j_M}_{j_i}{A}^{j_1\cdots j_M}_{j_i}$.
So there exists $O^{123}_{j_i}\in \mathrm{SO}(3)$,  $A'^{123}_{j_i}=O^{123}_{j_i}{A}^{123}_{j_i}$, which also holds for Eq.(\ref{T2X'2}) when $M=3$.

Combining with Eq.(\ref{T2X'2}), when $M<3,1\leq j_1<\cdots<j_M\leq 3$ we can get
\begin{equation*}
\begin{array}{ll}
(A'^{123}_{j_i})^{\mathrm{t}}{A'^{j_1\cdots j_M}_{j_i}}
&=({A}^{123}_{j_i})^{\mathrm{t}}(O^{123}_{j_i})^{\mathrm{t}}(O^{j_1\cdots j_M}_{j_i}{A}^{j_1\cdots j_M}_{j_i})\\
&=(A^{123}_{j_i})^{\mathrm{t}}{A}^{j_1\cdots j_M}_{j_i},
\end{array}
\end{equation*}
i.e., $(O^{123}_{j_i})^{\mathrm{t}}O^{j_1\cdots j_M}_{j_i}=I$.

With $O^{j_1\cdots j_M}_{j_i}\in\mathrm{SO}(3)$, we know that $O^{j_1\cdots j_M}_{j_i}=O^{123}_{j_i}$and let $O^{j_1\cdots j_M}_{j_i}=O_{j_i}$. That is to say, there are the same $O_1,O_2,O_3$, such that
${A}'^{j_1\cdots j_M}_{j_1}=O_{j_1}{A}^{j_1\cdots j_M}_{j_1},  \quad  \cdots, \quad {A}'^{j_1\cdots j_M}_{j_M}=O_{j_M}{A}^{j_1\cdots j_M}_{j_M} \ (M\leq 3,1\leq j_1<\cdots<j_M\leq 3$).

So $\rho$ is local unitary equivalent to $\rho'$ by Theorem \ref{t}.
$\hfill\Box$

More generally, Eq.(\ref{T2X'1}) and Eq.(\ref{T2X'2}) can be written in the following form (easily seen by Theorem \ref{h}):
$$(A^{j_1\cdots j_M}_{j_i})^{\mathrm{t}}A^{k_1\cdots k_N}_{k_l}=(A'^{j_1\cdots j_M}_{j_i})^{\mathrm{t}}{A}'^{k_1\cdots k_N}_{k_l}$$
where $j_{i}=k_{l}$, $i=1,\cdots,M; \ l=1,\cdots,N$, and
$M\leq 3$, $1\leq j_1<\cdots<j_M\leq 3$, and $N\leq 3$, $1\leq k_1<\cdots<k_N\leq 3$. 

The above results of necessary and sufficient conditions provide a method to detect whether two 3-qubit states are unitary equivalent.
One first
uses the factor matrices of tensors $\mathcal{X}_{123}$ and $\mathcal{X}'_{123}$ to find $O_1,O_2,O_3\in SO(3)$,
then one examines whether the first-order tensors $\mathcal{X}_{j_1}, \mathcal{X}'_{j_1}(j_1=1,2,3)$ and the second-order tensors $\mathcal{X}_{k_1k_2}, \mathcal{X}'_{k_1k_2}(1\leq k_1<k_2\leq 3)$ satisfy \eqref{lemma 2} with the same orthogonal matrices $O_{j_1}$.

Alternatively, if we can use the CP decompositions as follows. Note that the CP decomposition of $\mathcal{X}_{j_1}$ and $\mathcal{X}'_{j_1}$ are themselves. The CP decomposition of the matrices $\mathcal{X}_{k_1k_2}$ and $\mathcal{X}'_{k_1k_2}$ are easy and
are used to find appropriate factor matrices $A^{k_1k_2}_{k_1}$, $A^{k_1k_2}_{k_2}$ and $A'^{k_1k_2}_{k_1}$, $A'^{k_1k_2}_{k_2}$ satisfying
 Eq.(\ref{T2X'1}). Then, combining with vectors $A^{j_1}=\mathcal{X}_{j_1}$ and $A'^{j_1}=\mathcal{X}'_{j_1}$, we may choice suitable matrices $O_1,O_2,O_3\in SO(3)$ that make $A'^{k_1k_2}_{k_1}=O_{k_1}A^{k_1k_2}_{k_1}$, $A'^{k_1k_2}_{k_2}=O_{k_2}A^{k_1k_2}_{k_2}$, and $A'^{j_1}_{j_1}=O_{j_1}A^{j_1}_{j_1}(j_1=1,2,3)$. Finally, we just need to verify whether  $\mathcal{X}'_{123}=(O_1\otimes O_2\otimes O_3)\mathcal{X}_{123}$ is true.

The following part mainly gives some other conclusion about third-order tensor of format $4\times4\times4$.

Let
\begin{equation}\label{aet}
\mathcal{X}=(\tilde{x}_{i_1i_2i_3})
\end{equation}
represents the coefficient tensor
of $\rho$ in Eq.({\ref{rho}}), where $\tilde{x}_{i_1i_2i_3}=x_{i_1-1i_2-1i_3-1}(i_1,i_2,i_3=1,2,3,4)$. That is to say, using all coefficients of the state to construct only one third-order tensor.

\begin{remark}\label{n3}
 If two generic 3-qubit states are local unitary equivalent, $[\![A, B, C]\!]$ is the factor matrix of $\mathcal{X}$, there is factor matrix $[\![ A', B', C']\!]$ of $\mathcal{X}'$ and $O_k\in \mathrm{SO}(4)$ such that
\begin{equation}\label{4leq}
{A}'=O_1{A}, \quad {B}'=O_2{B}, \quad {C}'=O_3{C}.
\end{equation}
\end{remark}


The remark is clear from Lemma \ref{e}. Similarly one can get a necessary condition as Theorem \ref{h} for the $4\times4\times4$ tensor 
by $A^{\mathrm{t}}_{k}A_{k}=A'^{\mathrm{t}}_{k}A'_{k},\ k=1,2,3$, where $A_{k}$ and $A'_{k}$ denote the factor matrices of the tensor $\mathcal{\rho}$ and $\rho'$, respectively.

Especially, when the factor matrices are nonsingular square matrices, we can determine if two states are not local unitary equivalent by checking the correlation values $A_{k}'A_{k}^{-1}(k=1,2,3)$ of factor matrices of each CP decomposition for two states is not $SO(4)$ operator.

In the following we are going to study invariants of LU equivalence 3-qubit states using the CP decomposition. 

\begin{corollary}\label{coroll} Let $S$ be the set of local unitary equivalent
3-qubit states $\rho$ with the face matrices $[\![A, B, C]\!]$ 
such that $k_{\hat{A}}+k_{\hat{B}}+k_{\hat{C}}\geq 2R+2$
for their CP decompositions. Then any member of $S$ has the same values of the following
quantities:
\begin{eqnarray}\label{3geq}
&&(i) \ \ R, \ \lambda_r,\ k_{\hat{A}}, \ k_{\hat{B}}, \ k_{\hat{C}}, \\
\label{32geq}&&(ii) \ \ \mathrm{Tr}({\hat{A}}^{\mathrm{t}}{\hat{A}}), \ \mathrm{Tr}({\hat{B}}^{\mathrm{t}}{\hat{B}}), \ \mathrm{Tr}({\hat{C}}^{\mathrm{t}}{\hat{C}}),
\end{eqnarray}
where $R$ is the rank of the coefficient tensor $\mathcal{X}$ and $r=1,\cdots,R$.
\end{corollary}

\noindent
{\bf{Proof:}} It is clear from the proof of Lemma \ref{mm} that the rank is an invariant of $S$.
Let $\rho$ and $\rho'$ be two $3$-states belonging to $S$. Suppose their CP decompositions are
$\mathcal{X}=\sum^{R}_{r=1}\lambda_{r}{\hat{\mathbf{a}}}_{r}\circ{\hat{\mathbf{b}}}_{r}\circ{\hat{\mathbf{c}}}_{r}=[\![\mathbf{\lambda};{\hat{A}},{\hat{B}},{\hat{C}}]\!]$ and $\mathcal{X}'=\sum^{R}_{s=1}\lambda'_{s}{\hat{\mathbf{a}}'}_{s}\circ{\hat{\mathbf{b}}'}_{s}\circ{\hat{\mathbf{c}}'}_{s}=[\![\mathbf{\lambda'};{\hat{A}'},{\hat{B}'},{\hat{C}'}]\!]$, respectively. It follows from Remark \ref{n3} there exist orthogonal matrices $O_i\in SO(4)$ such that
\begin{equation}\label{Pth3}
\hat{A}'=O_1\hat{A}, \ \hat{B}'=O_2\hat{B}, \ \hat{C}'=O_3\hat{C}.
\end{equation}
It follows that  $k_{\hat{A}'}=k_{\hat{A}}$, $k_{\hat{B}'}=k_{\hat{B}}$, and $k_{\hat{C}'}=k_{\hat{C}}$.
Consequently the trace identities hold.


$\hfill\Box$

Recall that the mode-$i$ matrizations $X_{(i)}(i=1,2,3)$ of the tensor $\mathcal X$ associated with the state $\rho$ are defined as certain foldings.
We can see how they are changed under local unitary equivalence.

\begin{proposition}\label{xi}
Let $\rho$ be a
generic 3-qubit state with coefficient tensor $\mathcal X$. Under the local unitary equivalent the following Frobenius norms are invariant
\be\label{T22X}
\|{X}_{(1)}\|_{\mathrm{F}}, \ \|{X}_{(2)}\|_{\mathrm{F}}, \ \|{X}_{(3)}\|_{\mathrm{F}}.
\ee
Moreover the singular values of the matrizations $X_{(i)}$ are also invariant under the local unitary equivalence.
\end{proposition}

\noindent
{\bf{Proof:}} Suppose $\rho$ has a CP decomposition: $\mathcal{X}=\sum^{R}_{r=1}{\mathbf{a}}_{r}\circ{\mathbf{b}}_{r}\circ{\mathbf{c}}_{r}=[\![{A},{B},{C}]\!]$,
then $X_{(1)}=A(C\odot B)^{\mathrm{t}}$ etc. If $\rho'$ is local unitary equivalent to $\rho$, then there are orthogonal matrices $O_i$ such that the corresponding
$X'_{(1)}=(O_1A)(O_3C\odot O_2B)^t$. Therefore
\begin{equation*}
\begin{array}{ll}
\|{X}'_{(1)}\|_{\mathrm{F}}
&=\sqrt{\mathrm{Tr}(O_1{A}(({C}^{\mathrm{t}}O^{\mathrm{t}}_{3}O_{3}{C})\ast({B}^{\mathrm{t}}O^{\mathrm{t}}_{2}O_{2}{B})){A}^{\mathrm{t}}O^{\mathrm{t}}_{1})} \\
&=\sqrt{\mathrm{Tr}({A}({C}^{\mathrm{t}}{C}\ast{B}^{\mathrm{t}}{B}){A}^{\mathrm{t}})}
\\
&=\sqrt{\mathrm{Tr}(({A}({C}\odot{B})^{\mathrm{t}})({A}({C}\odot{B})^{\mathrm{t}})^{\mathrm{t}})} \\
&=\|{X}_{(1)}\|_{\mathrm{F}}.
\end{array}
\end{equation*}
Similarly we can show the other norms are invariant.
Note that same idea can easily show that the characteristic polynomial of $X_{(i)}X_{(i)}^t$ stays the same under the LU since
$O_i$ have determinant one, therefore the singular values of $X_{(i)}$
are also invariant.
$\hfill\Box$

When $\mathcal{X}\in\mathbb{R}^{4\times4\times4}$, the norm
$\|\mathcal{X}\|=\sqrt{\sum\limits^4_{i_1=1}\sum\limits^4_{i_2=1}\sum\limits^4_{i_3=1}\tilde{x}^{2}_{i_1i_2i_3}}$. As
 the entries of $X_{(1)}$ (or $X_{(2)}, \ X_{(3)}$) are also those of $\mathcal{X}$, $\|\mathcal{X}\|=\|{X}_{(i)}\|_{\mathrm{F}}, \ i=1,2,3$. The following is then clear.
\begin{corollary}\label{x1}
If two generic 3-qubit states are local unitary equivalent, their corresponding tensors $\mathcal{X}$ and $\mathcal{X}'$ have the same norms.
\end{corollary}

We need the following technical lemma to proceed.
{\begin{lemma}\label{ss}
Let $S\in \mathbb{R}^{\alpha\times\beta},\ N\in\mathbb{R}^{\gamma\times\delta},\ P\in\mathbb{R}^{\beta\times r}$, and $Q\in\mathbb{R}^{\delta\times r}$ be matrices of
said sizes, then
\begin{equation}\label{sf}
(S\otimes N)(P\odot Q)=(SP)\odot (NQ).
\end{equation}
where $\odot$ denotes the Khatri-Rao product.
\end{lemma}

This can be easily verified by writing $S, N$ as row blocks and $P, Q$ as column blocks.
\begin{corollary}\label{Xx}
If two generic 3-qubit states $\rho$ and $\rho'$ are local unitary equivalent, there exist $O_1,O_2$ and $O_3\in SO(4)$, such that
\begin{eqnarray}\label{coro}
&&X'_{(1)}=O_1X_{(1)}(O_3\otimes O_2)^{\mathrm{t}}, \\
&&X'_{(2)}=O_2X_{(2)}(O_3\otimes O_1)^{\mathrm{t}}, \\
&&X'_{(3)}=O_3X_{(3)}(O_2\otimes O_1)^{\mathrm{t}},
\end{eqnarray}
where $X_{(1)},X_{(2)},X_{(3)}$ and $X'_{(1)},X'_{(2)},X'_{(3)}$ are the mode-$n$ matricizations of the corresponding tensors of $\rho$ and $\rho'$ respectively.
\end{corollary}

Using the CP decomposition of the tensor, Lemma \ref{ss} and Eq.(\ref{4leq}) of Remark \ref{n3}, we immediately see the results.

\section{LU invariants of generic multi-qudit states}

{Let $\mathcal{X}=(x_{i_1i_2\cdots i_N})\in\mathbb{R}^{k_1\times k_2\times\cdots\times k_{N}}$ be an $N$-way tensor with rank $R$, then its CP decomposition is
\begin{equation}\label{NX}
\mathcal{X}=\sum^{R}_{r=1}\mathbf{a}^{(1)}_{r}\circ\mathbf{a}^{(2)}_{r}\circ\cdots\circ\mathbf{a}^{(N)}_{r}=[\![{A_{1}},{A_{2}},\cdots,{A_{N}}]\!],
\end{equation}
where $A_i=[\mathbf{a}_1^{(i)}, \cdots,  \mathbf{a}_R^{(i)}]\in\mathbb{R}^{k_n\times R}$.
A sufficient condition for essential uniqueness is
\begin{equation}\label{Nsc}
\sum^{N}_{n=1}k_{A_{n}}\geq2R+(N-1).
\end{equation}
In this case, the mode-$n$ matricization \cite{Tencp} of $\mathcal{X}$ is given by
\begin{equation}\label{nmatricization}
    X_{(n)}=A_{n}(A_{N}\odot\cdots\odot A_{n+1}\odot A_{n-1}\odot \cdots\odot A_{1})^\mathrm{t}.
\end{equation}



Given a multi-qudit state $\rho\in H^{d_1}_1\otimes H^{d_2}_2\otimes\cdots\otimes H^{d_N}_{N}$, where $H_k\simeq\mathbb{C}^{d_k}$ are the Hilbert spaces of subsystems $k (1\leq k\leq N)$, then
\begin{equation}\label{mrho}
\rho=\sum^{d^2_1}_{i_1=0}\sum^{d^2_2}_{i_2=0}\cdots\sum^{d^2_N}_{i_N=0}x_{i_1i_2\cdots i_N}\lambda^{(1)}_{i_1}\otimes\lambda^{(2)}_{i_2}\otimes\cdots\otimes\lambda^{(N)}_{i_N},
\end{equation}
where $\lambda^{(k)}_0=\mathrm{I}_{d_k}$ is the identity, $k=1,\cdots,N$, and $\lambda^{(k)}_{i_k}$ is the traceless Hermitian generators of the group $SU(d_k),k=1,2,\cdots,N,i_k=1,2,\cdots,d^2_k-1$.

The generators ${\lambda_{k}},k=1,2,\cdots,d_{k}^{2}-1$ of $SU(d_k)$ can be constructed from any orthonormal basis in $H_k$;
when $k=1,2,\cdots,d_k-1$,
$$\lambda_k=\omega_{l}\equiv\sqrt{\frac{2}{(l+1)(l+2)}}(\sum^{l}_{j=0}|j\rangle\langle j|-(l+1)|l+1\rangle\langle l+1|);$$
when $k=d_k,\cdots,(d_k+2)(d_k-1)/2$,
$$\lambda_k=u_{mn}\equiv|m\rangle\langle n|+|n\rangle\langle m|;$$
when $k=d_k(d_k+1)/2,\cdots,d_k^2-1$,
$$\lambda_k=v_{mn}\equiv-\sqrt{-1}(|m\rangle\langle n|-|n\rangle\langle m|),$$ where $0\leq l\leq d_k-2$, $0\leq m<n\leq d_k-1$, 
and
\begin{equation}\label{delta definition}
\mathrm{Tr}(\lambda_{i}\lambda_{j})=2\delta_{ij}=\left\{
\begin{aligned}
 2, \ & i=j \\
 0, \ & i\neq j
\end{aligned}
\right..
\end{equation}

Therefore
\begin{equation}\label{xxx}
x_{i_1i_2\cdots i_N}={\frac{1}{2^{N-M}\cdot d_{k_0}d_{k_1}\cdots d_{k_M}}}\mathrm{Tr}(\rho\lambda^{(1)}_{i_1-1}\otimes\lambda^{(2)}_{i_2-1}\otimes\cdots\lambda^{(N)}_{i_N-1}),
\end{equation}
where there are exactly $M$-superscripts equal to $1$: $i_{k_{1}}=i_{k_{2}}=\cdots=i_{k_{M}}=1$, and $d_{k_0}=1$, $M\in \{0,1,\cdots,N\}$, $k_1,k_2,\cdots,k_M\in \{1,2,\cdots,N\}$.
The following is clear.
%
%
%

\begin{lemma}\label{a} If two mixed states $\rho$ and $\rho'$ are local unitary equivalent, then there are real orthogonal matrices $O_{k}\in SO(d_k^2),k=1,\cdots,N$ such that
\be\label{lemma 4}
(O_1\otimes O_2\otimes\cdots\otimes O_N)\mathcal{X}=\mathcal{X}' ,
\ee
where $\mathcal{X}=(\tilde{x}_{i_1\cdots i_N})$ and $\mathcal{X}'=(\tilde{x}'_{i_1\cdots i_N})$ are the coefficient tensors of $\rho$ and $\rho'$ respectively.
\end{lemma}

{When $d_k=2$ for all $k$, let $\mathcal{X}_{j_1j_2\cdots j_M}(M\leq N,1\leq j_1<j_2<\cdots<j_M\leq N)$ be the $M$th-order tensor defined by Def. \ref{xii}.
The necessary and sufficient condition for the local unitary equivalent about two N-qubit states are obtained by the method similar to the case of 3-qubit state.



\begin{theorem}\label{Nth} Suppose the coefficient tensor $\mathcal{X}$ of the generic N-qudit state $\rho$ has
the CP decomposition:
$\mathcal{X}=\sum^{R}_{r=1}\mathbf{a}^{(1)}_{r}\circ\mathbf{a}^{(2)}_{r}\circ\cdots\circ\mathbf{a}^{(N)}_{r}=[\![{A_1},{A_2},\cdots,{A_N}]\!]
$. If $\rho'$ is local unitary equivalent to $\rho$, then the coefficient tensor $\mathcal{X}'$ has a CP decomposition such that
\begin{eqnarray}\label{Nleq}
&&(i) \ \ \mathcal{X}'=\sum^{R}_{r=1}\mathbf{a}^{'(1)}_{r}\circ\mathbf{a}^{'(2)}_{r}\circ\cdots\circ\mathbf{a}^{'(N)}_{r}=[\![{A_1}',{A_2}',\cdots,{A_N}']\!],
\\
&&(ii) \ \ {A}'_{k}=O_k{A_k}, \\
&&(iii) \ \ {A}^{'\mathrm{t}}_{k}A'_{k}=A^{\mathrm{t}}_kA_k,
\end{eqnarray}
where $O_k\in SO(d_k^2)$. 
\end{theorem}


The above result is proved similarly as in the $3$-qubit case, and we only need to
note that when two states are local unitary equivalent, it follows from Lemma \ref{a} that $\mathcal{X}'=(O_1\otimes\cdots\otimes O_N)\mathcal{X}=(O_1\otimes\cdots\otimes O_N)\sum^{R}_{r=1}\mathbf{a}^{(1)}_{r}\circ\mathbf{a}^{(2)}_{r}\circ\cdots\circ\mathbf{a}^{(N)}_{r}
=\sum^{R}_{r=1}(O_{1}\mathbf{a}^{(1)}_{r})\circ \cdots\circ (O_{N}\mathbf{a}^{(N)}_{r})=[\![O_{1}{A_1},\cdots,O_{N}{A_N}]\!]$, then the result is clear.

When the factor matrices of $\rho$ are normalized with unit columns: $\mathcal{X}=\sum^{R}_{r=1}\lambda_r\mathbf{\hat{a}}^{(1)}_{r}\circ\mathbf{\hat{a}}^{(2)}_{r}\circ\cdots\circ\mathbf{\hat{a}}^{(N)}_{r}=[\![\mathbf{\lambda};\hat{A}_1,\hat{A}_2,\cdots\hat{A}_N]\!]$
and assume that $\sum^{N}_{n=1}k_{\hat{A}_{n}}\geq2R+(N-1)$, then the existence of the CP
decomposition is guaranteed.

\begin{theorem}\label{Neo}
Let $S$ be the set of local unitary equivalent
N-qubit states $\rho$ with the face matrices $[\![X_{(1)}, \cdots , X_{(N)}]\!]$ 
for their CP decompositions. Then any member of $S$ has the same values of the following
quantities:
\begin{eqnarray}\label{Th5}
&&(i) \ \ \|{X}_{(k)}\|_{\mathrm{F}}, \ \|\mathcal{X}\|, \\
&&(ii) \ \ the \ singular \ values \ of \ {X}_{(k)},
\end{eqnarray}
where $X_{(k)}$ is the mode-$n$ matricization of the tensor $\mathcal{X}$, $\|\cdot\|_{\mathrm{F}}$ stands for the Frobenius norm and $k=1,\cdots,N$.
\end{theorem}
The theorem is easily shown by noting that
$X_{(k)}'=O_{k}X_{(k)}(O_{N}\otimes\cdots\otimes O_{k+1}\otimes O_{k-1}\cdots\otimes O_{1})^{\mathrm{t}}$ due to
Eq.(\ref{sf}) and Eq.(\ref{nmatricization}).

{\bf Example 1:} Consider the qutrit Werner state:
\begin{equation}\label{Ws}
\rho(z)=\frac{1-z}{9}\mathrm{I}+z|\psi\rangle\langle\psi|,
\end{equation}
where $|\psi\rangle=\frac{1}{\sqrt{3}}(|00\rangle+|11\rangle+|22\rangle)$ and $z\in[0,1]$.

(i) Let $\mathcal X$ be the coefficient tensor of $\rho(1)$. It is easy to see
\begin{equation*}
\begin{array}{ll}
\mathcal{X}=\left(\begin{smallmatrix}
     1/9 & 0 & 0 & 0 & 0 & 0 & 0 & 0 & 0 \\
     0 & 1/6 & 0 & 0 & 0 & 0 & 0 & 0 & 0 \\
     0 & 0 & 1/6 & 0 & 0 & 0 & 0 & 0 & 0 \\
     0 & 0 & 0 & 1/6 & 0 & 0 & 0 & 0 & 0 \\
     0 & 0 & 0 & 0 & 1/6 & 0 & 0 & 0 & 0 \\
     0 & 0 & 0 & 0 & 0 & 1/6 & 0 & 0 & 0 \\
     0 & 0 & 0 & 0 & 0 & 0 & -1/6 & 0 & 0 \\
     0 & 0 & 0 & 0 & 0 & 0 & 0 & -1/6 & 0 \\
     0 & 0 & 0 & 0 & 0 & 0 & 0 & 0 & -1/6
\end{smallmatrix}\right),
\end{array}
\end{equation*}
and the rank of $\mathcal{X}$ is $R=9$. Its CP decomposition $\mathcal{X}=AB^{\mathrm{T}}=\sum\limits^{9}_{r=1}\mathbf{a}_{r}\circ\mathbf{b}_{r}$, where
$A=\left(\begin{smallmatrix}
     0 & 0 & 0 & 0 & 0 & 0 & 0 & 0 & 1/9 \\
     0 & 1/6 & 0 & 0 & 0 & 0 & 0 & 0 & 0 \\
     0 & 0 & 1/6 & 0 & 0 & 0 & 0 & 0 & 0 \\
     0 & 0 & 0 & 1/6 & 0 & 0 & 0 & 0 & 0 \\
     0 & 0 & 0 & 0 & 1/6 & 0 & 0 & 0 & 0 \\
     0 & 0 & 0 & 0 & 0 & 1/6 & 0 & 0 & 0 \\
     0 & 0 & 0 & 0 & 0 & 0 & 1/6 & 0 & 0 \\
     0 & 0 & 0 & 0 & 0 & 0 & 0 & 1/6 & 0 \\
     1/6 & 0 & 0 & 0 & 0 & 0 & 0 & 0 & 0
\end{smallmatrix}\right)$,
$B=\left(\begin{smallmatrix}
     0 & 0 & 0 & 0 & 0 & 0 & 0 & 0 & 1 \\
     0 & 1 & 0 & 0 & 0 & 0 & 0 & 0 & 0 \\
     0 & 0 & 1 & 0 & 0 & 0 & 0 & 0 & 0 \\
     0 & 0 & 0 & 1 & 0 & 0 & 0 & 0 & 0 \\
     0 & 0 & 0 & 0 & 1 & 0 & 0 & 0 & 0 \\
     0 & 0 & 0 & 0 & 0 & 1 & 0 & 0 & 0 \\
     0 & 0 & 0 & 0 & 0 & 0 & -1 & 0 & 0 \\
     0 & 0 & 0 & 0 & 0 & 0 & 0 & -1 & 0 \\
     -1 & 0 & 0 & 0 & 0 & 0 & 0 & 0 & 0
\end{smallmatrix}\right).$

Now consider the state $\rho'$ with the tensor coefficient
\begin{equation*}
\begin{array}{ll}
\mathcal{X}'=\left(\begin{smallmatrix}
     1/9 & 0 & 0 & 0 & 0 & 0 & 0 & 0 & 0 \\
     0 & -1/12 & -\sqrt{3}/12 & 0 & 0 & 0 & 0 & 0 & 0 \\
     0 & \sqrt{3}/12 & -1/12 & 0 & 0 & 0 & 0 & 0 & 0 \\
     0 & 0 & 0 & 0 & 1/6 & 0 & 0 & 0 & 0 \\
     0 & 0 & 0 & 0 & 0 & 1/6 & 0 & 0 & 0 \\
     0 & 0 & 0 & 1/6 & 0 & 0 & 0 & 0 & 0 \\
     0 & 0 & 0 & 0 & 0 & 0 & 0 & 1/6 & 0 \\
     0 & 0 & 0 & 0 & 0 & 0 & 0 & 0 & 1/6 \\
     0 & 0 & 0 & 0 & 0 & 0 & -1/6 & 0 & 0
\end{smallmatrix}\right),
\end{array}
\end{equation*}
with the CP decomposition $\mathcal{X}'=A'B'^{\mathrm{T}}=\sum\limits^{9}_{r=1}\mathbf{a}'_{r}\circ\mathbf{b}'_{r}$, where the factor matrices are given by
$A'=\left(\begin{smallmatrix}
     0 & 0 & 0 & 0 & 0 & 0 & 0 & 0 & 1/9 \\
     0 & -1/12 & \sqrt{3}/12 & 0 & 0 & 0 & 0 & 0 & 0 \\
     0 & \sqrt{3}/12 & 1/12 & 0 & 0 & 0 & 0 & 0 & 0 \\
     0 & 0 & 0 & 0 & -1/6 & 0 & 0 & 0 & 0 \\
     0 & 0 & 0 & 0 & 0 & 1/6 & 0 & 0 & 0 \\
     0 & 0 & 0 & -1/6 & 0 & 0 & 0 & 0 & 0 \\
     0 & 0 & 0 & 0 & 0 & 0 & 0 & -1/6 & 0 \\
     -1/6 & 0 & 0 & 0 & 0 & 0 & 0 & 0 & 0 \\
     0 & 0 & 0 & 0 & 0 & 0 & 1/6 & 0 & 0
\end{smallmatrix}\right)$,
$B'=\left(\begin{smallmatrix}
     0 & 0 & 0 & 0 & 0 & 0 & 0 & 0 & 1 \\
     0 & 1 & 0 & 0 & 0 & 0 & 0 & 0 & 0 \\
     0 & 0 & -1 & 0 & 0 & 0 & 0 & 0 & 0 \\
     0 & 0 & 0 & -1 & 0 & 0 & 0 & 0 & 0 \\
     0 & 0 & 0 & 0 & -1 & 0 & 0 & 0 & 0 \\
     0 & 0 & 0 & 0 & 0 & 1 & 0 & 0 & 0 \\
     0 & 0 & 0 & 0 & 0 & 0 & -1 & 0 & 0 \\
     0 & 0 & 0 & 0 & 0 & 0 & 0 & -1 & 0 \\
     -1 & 0 & 0 & 0 & 0 & 0 & 0 & 0 & 0
\end{smallmatrix}\right)$.

One sees that there exist $O_1,O_2\in SO(9)$ such that $A'=O_1A$, $B'=O_2B$, i.e.,
\begin{equation}
\begin{array}{ll}
O_1=\left(\begin{smallmatrix}
     1 & 0 & 0 & 0 & 0 & 0 & 0 & 0 & 0 \\
     0 & -1/2 & \sqrt{3}/2 & 0 & 0 & 0 & 0 & 0 & 0 \\
     0 & \sqrt{3}/2 & 1/2 & 0 & 0 & 0 & 0 & 0 & 0 \\
     0 & 0 & 0 & 0 & -1 & 0 & 0 & 0 & 0 \\
     0 & 0 & 0 & 0 & 0 & 1 & 0 & 0 & 0 \\
     0 & 0 & 0 & -1 & 0 & 0 & 0 & 0 & 0 \\
     0 & 0 & 0 & 0 & 0 & 0 & 0 & -1 & 0 \\
     0 & 0 & 0 & 0 & 0 & 0 & 0 & 0 & -1 \\
     0 & 0 & 0 & 0 & 0 & 0 & 1 & 0 & 0
\end{smallmatrix}\right)
\end{array}, \
\begin{array}{ll}
O_2=\left(\begin{smallmatrix}
     1 & 0 & 0 & 0 & 0 & 0 & 0 & 0 & 0 \\
     0 & 1 & 0 & 0 & 0 & 0 & 0 & 0 & 0 \\
     0 & 0 & -1 & 0 & 0 & 0 & 0 & 0 & 0 \\
     0 & 0 & 0 & -1 & 0 & 0 & 0 & 0 & 0 \\
     0 & 0 & 0 & 0 & -1 & 0 & 0 & 0 & 0 \\
     0 & 0 & 0 & 0 & 0 & 1 & 0 & 0 & 0 \\
     0 & 0 & 0 & 0 & 0 & 0 & 1 & 0 & 0 \\
     0 & 0 & 0 & 0 & 0 & 0 & 0 & 1 & 0 \\
     0 & 0 & 0 & 0 & 0 & 0 & 0 & 0 & 1
\end{smallmatrix}\right),
\end{array}
\end{equation}
therefore $\rho\overset{LOU}\simeq \rho'$ . In fact, $\rho'=(U_1\otimes U_2)\rho(U_1\otimes U_2)^{\dagger}$,
with $U_1=\left(\begin{smallmatrix}
     0 & 1 & 0 \\
     0 & 0 & 1 \\
     1 & 0 & 0
\end{smallmatrix}\right)$ and $U_2=\left(\begin{smallmatrix}
     0 & 0 & 1 \\
     1 & 0 & 0 \\
     0 & 1 & 0
\end{smallmatrix}\right)$.

(ii) Now let $\mathcal{Y}$ be the coefficient tensor of
$\rho(1/4)$:
\begin{equation*}
\begin{array}{ll}
\mathcal{Y}=\left(\begin{smallmatrix}
     1/9 & 0 & 0 & 0 & 0 & 0 & 0 & 0 & 0 \\
     0 & 1/24 & 0 & 0 & 0 & 0 & 0 & 0 & 0 \\
     0 & 0 & 1/24 & 0 & 0 & 0 & 0 & 0 & 0 \\
     0 & 0 & 0 & 1/24 & 0 & 0 & 0 & 0 & 0 \\
     0 & 0 & 0 & 0 & 1/24 & 0 & 0 & 0 & 0 \\
     0 & 0 & 0 & 0 & 0 & 1/24 & 0 & 0 & 0 \\
     0 & 0 & 0 & 0 & 0 & 0 & -1/24 & 0 & 0 \\
     0 & 0 & 0 & 0 & 0 & 0 & 0 & -1/24 & 0 \\
     0 & 0 & 0 & 0 & 0 & 0 & 0 & 0 & -1/24
\end{smallmatrix}\right).
\end{array}
\end{equation*}
Clearly $\sqrt{34}/36=\|\mathcal{Y}\|\neq\|\mathcal{X}\|=\sqrt{19}/9$, so the two states $\rho(1)$ and $\rho(1/4)$ are not local unitary equivalent.

{\bf Example 2:}  Consider two mixed 3-qubit quantum states: $\rho=\frac{2}{17}(2|\psi_{+}\rangle\langle\psi_{+}|+|001\rangle\langle001|+|010\rangle\langle010|+2|011\rangle\langle011|
+\frac{1}{2}|100\rangle\langle100|+|101\rangle\langle101|+|110\rangle\langle110|)$ with $|\psi_{+}\rangle=\frac{1}{\sqrt{2}}(|000\rangle+|111\rangle)$ and
\begin{equation}\label{exam2}
\tau=\frac{2}{17}\left(\begin{smallmatrix}
     1 & 0 & 0 & 0 & 0 & \frac{1}{2} & 0 & \frac{1}{2} \\
     0 & \frac{3}{2} & 0 & \frac{1}{2} & 0 & 0 & 0 & 0 \\
     0 & 0 & 1 & 0 & 0 & -\frac{1}{2} & 0 & -\frac{1}{2} \\
     0 & \frac{1}{2} & 0 & \frac{3}{2} & 0 & 0 & 0 & 0 \\
     0 & 0 & 0 & 0 & \frac{3}{4} & 0 & \frac{1}{4} & 0 \\
     \frac{1}{2} & 0 & -\frac{1}{2} & 0 & 0 & 1 & 0 & 0 \\
     0 & 0 & 0 & 0 & \frac{1}{4} & 0 & \frac{3}{4} & 0 \\
     \frac{1}{2} & 0 & -\frac{1}{2} & 0 & 0 & 0 & 0 & 1
\end{smallmatrix}\right).
\end{equation}
The coefficient tensor $\rho$ has the frontal slices:
$X_1=\left(\begin{smallmatrix}
     1/8 & 0 & 0 & -3/136 \\
     0 & 0 & 0 & 0 \\
     0 & 0 & 0 & 0 \\
     3/136 & 0 & 0 & -1/136
\end{smallmatrix}\right), \
X_2=\left(\begin{smallmatrix}
     0 & 0 & 0 & 0 \\
     0 & 1/34 & 0 & 0 \\
     0 & 0 & -1/34 & 0 \\
     0 & 0 & 0 & 0
\end{smallmatrix}\right), \
X_3=\left(\begin{smallmatrix}
     0 & 0 & 0 & 0 \\
     0 & 0 & -1/34 & 0 \\
     0 & -1/34 & 0 & 0 \\
     0 & 0 & 0 & 0
\end{smallmatrix}\right)$, and
$X_4=\left(\begin{smallmatrix}
     -3/136 & 0 & 0 & 1/136 \\
     0 & 0 & 0 & 0 \\
     0 & 0 & 0 & 0 \\
     -1/136 & 0 & 0 & 3/136
\end{smallmatrix}\right)$.
The three mode-$n$ unfoldings are
\begin{equation}
\begin{array}{ll}
X_{(1)}=\left(\begin{smallmatrix}
     1/8 & 0 & 0 & -3/136 & 0 & 0 & 0 & 0 & 0 & 0 & 0 & 0 & -3/136 & 0 & 0 & 1/136\\
     0 & 0 & 0 & 0 & 0 & 1/34 & 0 & 0 & 0 & 0 & -1/34 & 0 & 0 & 0 & 0 & 0\\
     0 & 0 & 0 & 0 & 0 & 0 & -1/34 & 0 & 0 & -1/34 & 0 & 0 & 0 & 0 & 0 & 0\\
     3/136 & 0 & 0 & -1/136 & 0 & 0 & 0 & 0 & 0 & 0 & 0 & 0 & -1/136 & 0 & 0 & 3/136
\end{smallmatrix}\right)
\end{array},
\end{equation}
\begin{equation}
\begin{array}{ll}
X_{(2)}=\left(\begin{smallmatrix}
     1/8 & 0 & 0 & 3/136 & 0 & 0 & 0 & 0 & 0 & 0 & 0 & 0 & -3/136 & 0 & 0 & -1/136\\
     0 & 0 & 0 & 0 & 0 & 1/34 & 0 & 0 & 0 & 0 & -1/34 & 0 & 0 & 0 & 0 & 0\\
     0 & 0 & 0 & 0 & 0 & 0 & -1/34 & 0 & 0 & -1/34 & 0 & 0 & 0 & 0 & 0 & 0\\
     -3/136 & 0 & 0 & -1/136 & 0 & 0 & 0 & 0 & 0 & 0 & 0 & 0 & 1/136 & 0 & 0 & 3/136
\end{smallmatrix}\right),
\end{array}
\end{equation}
\begin{equation}
\begin{array}{ll}
X_{(3)}=\left(\begin{smallmatrix}
     1/8 & 0 & 0 & 3/136 & 0 & 0 & 0 & 0 & 0 & 0 & 0 & 0 & -3/136 & 0 & 0 & -1/136\\
     0 & 0 & 0 & 0 & 0 & 1/34 & 0 & 0 & 0 & 0 & -1/34 & 0 & 0 & 0 & 0 & 0\\
     0 & 0 & 0 & 0 & 0 & 0 & -1/34 & 0 & 0 & -1/34 & 0 & 0 & 0 & 0 & 0 & 0\\
     -3/136 & 0 & 0 & -1/136 & 0 & 0 & 0 & 0 & 0 & 0 & 0 & 0 & 1/136 & 0 & 0 & 3/136
\end{smallmatrix}\right).
\end{array}
\end{equation}

The frontal slices of the coefficient tensor $\mathcal{X}'$ for $\tau$ are
$X'_1=\left(\begin{smallmatrix}
     1/8 & 3/136 & 0 & 0 \\
     0 & 0 & 0 & 0 \\
     0 & 0 & 0 & 0 \\
     3/136 & 1/136 & 0 & 0
\end{smallmatrix}\right), \
X'_2=\left(\begin{smallmatrix}
     0 & 0 & 0 & 0 \\
     0 & 0 & 0 & 1/34 \\
     0 & 0 & -1/34 & 0 \\
     0 & 0 & 0 & 0
\end{smallmatrix}\right), \
X'_3=\left(\begin{smallmatrix}
     0 & 0 & 0 & 0 \\
     0 & 0 & -1/34 & 0 \\
     0 & 0 & 0 & -1/34 \\
     0 & 0 & 0 & 0
\end{smallmatrix}\right)$, and
$X'_4=\left(\begin{smallmatrix}
     -3/136 & -1/136 & 0 & 0 \\
     0 & 0 & 0 & 0 \\
     0 & 0 & 0 & 0 \\
     -1/136 & -3/136 & 0 & 0
\end{smallmatrix}\right)$.
The three mode-$n$ unfoldings are
\begin{equation}
\begin{array}{ll}
X'_{(1)}=\left(\begin{smallmatrix}
     1/8 & 3/136 & 0 & 0 & 0 & 0 & 0 & 0 & 0 & 0 & 0 & 0 & -3/136 & -1/136 & 0 & 0\\
     0 & 0 & 0 & 0 & 0 & 0 & 0 & 1/34 & 0 & 0 & -1/34 & 0 & 0 & 0 & 0 & 0\\
     0 & 0 & 0 & 0 & 0 & 0 & -1/34 & 0 & 0 & 0 & 0 & -1/34 & 0 & 0 & 0 & 0\\
     3/136 & 1/136 & 0 & 0 & 0 & 0 & 0 & 0 & 0 & 0 & 0 & 0 & -1/136 & -3/136 & 0 & 0
\end{smallmatrix}\right)
\end{array},
\end{equation}
\begin{equation}
\begin{array}{ll}
X'_{(2)}=\left(\begin{smallmatrix}
     1/8 & 0 & 0 & 3/136 & 0 & 0 & 0 & 0 & 0 & 0 & 0 & 0 & -3/136 & 0 & 0 & -1/136\\
     3/136 & 0 & 0 & 1/136 & 0 & 0 & 0 & 0 & 0 & 0 & 0 & 0 & -1/136 & 0 & 0 & -3/136\\
     0 & 0 & 0 & 0 & 0 & 0 & -1/34 & 0 & 0 & -1/34 & 0 & 0 & 0 & 0 & 0 & 0\\
     0 & 0 & 0 & 0 & 0 & 1/34 & 0 & 0 & 0 & 0 & -1/34 & 0 & 0 & 0 & 0 & 0
\end{smallmatrix}\right),
\end{array}
\end{equation}
\begin{equation}
\begin{array}{ll}
X_{(3)}=\left(\begin{smallmatrix}
     1/8 & 0 & 0 & 3/136 & 3/136 & 0 & 0 & 1/136 & 0 & 0 & 0 & 0 & 0 & 0 & 0 & 0\\
     0 & 0 & 0 & 0 & 0 & 0 & 0 & 0 & 0 & 0 & -1/34 & 0 & 0 & 1/34 & 0 & 0\\
     0 & 0 & 0 & 0 & 0 & 0 & 0 & 0 & 0 & -1/34 & 0 & 0 & 0 & 0 & -1/34 & 0\\
     -3/136 & 0 & 0 & -1/136 & -1/136 & 0 & 0 & -3/136 & 0 & 0 & 0 & 0 & 0 & 0 & 0 & 0
\end{smallmatrix}\right).
\end{array}
\end{equation}

Clearly $\|X_{(1)}\|=\|X'_{(1)}\|=\|X_{(2)}\|=\|X'_{(2)}\|=\|X_{(3)}\|=\|X'_{(3)}\|=\frac{7\sqrt{578}}{1156}$.
This prompts us to guess that the two states may be
local unitary equivalent. Now we use
Theorem \ref{t} to analyze. For the state $\rho$, the first-order tensors are
\begin{equation}\label{xa}
\begin{array}{ll}
\mathcal{X}_{1}=\left(\begin{smallmatrix}
     0\\
     0\\
     3/136
\end{smallmatrix}\right)
\end{array}, \
\begin{array}{ll}
\mathcal{X}_{2}=\left(\begin{smallmatrix}
     0 \\
     0 \\
     -3/136
\end{smallmatrix}\right)
\end{array}, \
\begin{array}{ll}
\mathcal{X}_{3}=\left(\begin{smallmatrix}
     0 \\
     0 \\
     -3/136
\end{smallmatrix}\right),
\end{array}
\end{equation}
the second-order tensors are
\begin{equation}\label{x123}
\begin{array}{ll}
\mathcal{X}_{12}=\left(\begin{smallmatrix}
     0 & 0 & 0\\
     0 & 0 & 0\\
     0 & 0 & -1/136
\end{smallmatrix}\right)
\end{array}, \
\begin{array}{ll}
\mathcal{X}_{13}=\left(\begin{smallmatrix}
     0 & 0 & 0\\
     0 & 0 & 0\\
     0 & 0 & -1/136
\end{smallmatrix}\right)
\end{array}, \
\begin{array}{ll}
\mathcal{X}_{23}=\left(\begin{smallmatrix}
     0 & 0 & 0\\
     0 & 0 & 0\\
     0 & 0 & 1/136
\end{smallmatrix}\right),
\end{array}
\end{equation}
where the factor matrices of $\mathcal{X}_{12}$, $\mathcal{X}_{13}$ and $\mathcal{X}_{23}$ are $A^{12}_1=\left(\begin{smallmatrix}
     0 & 0 & 0 \\
     0 & 0 & 0 \\
     1/136 & 0 & 0
\end{smallmatrix}\right)$, $A^{12}_2=\left(\begin{smallmatrix}
     0 & 0 & 1 \\
     0 & 1 & 0 \\
     -1 & 0 & 0
\end{smallmatrix}\right)$, $A^{13}_1=\left(\begin{smallmatrix}
     0 & 0 & 0 \\
     0 & 0 & 0 \\
     1/136 & 0 & 0
\end{smallmatrix}\right)$, $A^{13}_3=\left(\begin{smallmatrix}
     0 & 0 & 1 \\
     0 & 1 & 0 \\
     -1 & 0 & 0
\end{smallmatrix}\right)$, and $A^{23}_2=\left(\begin{smallmatrix}
     0 & 0 & 0 \\
     0 & 0 & 0 \\
     0 & 0 & 1/136
\end{smallmatrix}\right)$, $A^{23}_3=\mathrm{I}$, respectively.

The third-order tensor $\mathcal{X}_{123}=(x_{i_1i_2i_3})(i_j\neq 0,j=1,2,3)$, where $x_{111}=1/34,\ x_{122}=x_{212}=x_{221}=-1/34,\ x_{333}=3/136$ and the other elements are 0. Therefore,
\begin{equation}\label{xx123}
\begin{array}{ll}
&X^{123}_{(1)}=\left(\begin{smallmatrix}
     1/34 & 0 & 0 & 0 & -1/34 & 0 & 0 & 0 & 0 \\
     0 & -1/34 & 0 & -1/34 & 0 & 0 & 0 & 0 & 0 & \\
     0 & 0 & 0 & 0 & 0 & 0 & 0 & 0 & 3/136
\end{smallmatrix}\right),\
X^{123}_{(2)}=\left(\begin{smallmatrix}
     1/34 & 0 & 0 & 0 & -1/34 & 0 & 0 & 0 & 0 \\
     0 & -1/34 & 0 & -1/34 & 0 & 0 & 0 & 0 & 0 & \\
     0 & 0 & 0 & 0 & 0 & 0 & 0 & 0 & 3/136
\end{smallmatrix}\right), \\ \\
&X^{123}_{(3)}=\left(\begin{smallmatrix}
     1/34 & 0 & 0 & 0 & -1/34 & 0 & 0 & 0 & 0 \\
     0 & -1/34 & 0 & -1/34 & 0 & 0 & 0 & 0 & 0 & \\
     0 & 0 & 0 & 0 & 0 & 0 & 0 & 0 & 3/136
\end{smallmatrix}\right).
\end{array}
\end{equation}

For the state $\rho'$, the first-order tensors are
\begin{equation}\label{xa'}
\begin{array}{ll}
\mathcal{X}'_{1}=\left(\begin{smallmatrix}
     0\\
     0\\
     3/136
\end{smallmatrix}\right)
\end{array}, \
\begin{array}{ll}
\mathcal{X}'_{2}=\left(\begin{smallmatrix}
     3/136 \\
     0 \\
     0
\end{smallmatrix}\right)
\end{array}, \
\begin{array}{ll}
\mathcal{X}'_{3}=\left(\begin{smallmatrix}
     0 \\
     0 \\
     -3/136
\end{smallmatrix}\right),
\end{array}
\end{equation}
and the second-order tensors are
\begin{equation}\label{x'123}
\begin{array}{ll}
\mathcal{X}'_{12}=\left(\begin{smallmatrix}
     0 & 0 & 0\\
     0 & 0 & 0\\
     1/136 & 0 & 0
\end{smallmatrix}\right)
\end{array}, \
\begin{array}{ll}
\mathcal{X}'_{13}=\left(\begin{smallmatrix}
     0 & 0 & 0\\
     0 & 0 & 0\\
     0 & 0 & -1/136
\end{smallmatrix}\right)
\end{array}, \
\begin{array}{ll}
\mathcal{X}'_{23}=\left(\begin{smallmatrix}
     0 & 0 & -1/136\\
     0 & 0 & 0\\
     0 & 0 & 0
\end{smallmatrix}\right),
\end{array}
\end{equation}
where the factor matrices of $\mathcal{X}'_{12}$, $\mathcal{X}'_{13}$ and $\mathcal{X}'_{23}$ are $A'^{12}_1=\left(\begin{smallmatrix}
     0 & 0 & 0 \\
     0 & 0 & 0 \\
     1/136 & 0 & 0
\end{smallmatrix}\right)$, $A'^{12}_2=\mathrm{I}$, $A'^{13}_1=\left(\begin{smallmatrix}
     0 & 0 & 0 \\
     0 & 0 & 0 \\
     1/136 & 0 & 0
\end{smallmatrix}\right)$, $A'^{13}_3=\left(\begin{smallmatrix}
     0 & 0 & 1 \\
     0 & 1 & 0 \\
     -1 & 0 & 0
\end{smallmatrix}\right)$, and $A'^{23}_2=\left(\begin{smallmatrix}
     0 & 0 & -1/136 \\
     0 & 0 & 0 \\
     0 & 0 & 0
\end{smallmatrix}\right)$, $A'^{23}_3=\mathrm{I}$, respectively.

The third-order tensor $\mathcal{X}'_{123}=(x_{i_1i_2i_3})(i_j\neq 0,j=1,2,3)$, where $x_{131}=1/34,\ x_{122}=x_{221}=x_{232}=-1/34,\ x_{313}=-3/136$ and the other elements are 0. In the same way, we know
\begin{equation}\label{xx'123}
\begin{array}{ll}
&X'^{123}_{(1)}=\left(\begin{smallmatrix}
      0& 0 & 1/34 & 0 & -1/34 & 0 & 0 & 0 & 0 \\
     0 & -1/34 & 0 & 0 & 0 & -1/34 & 0 & 0 & 0 & \\
     0 & 0 & 0 & 0 & 0 & 0 & 0 & 0 & 3/136
\end{smallmatrix}\right),\
X'^{123}_{(2)}=\left(\begin{smallmatrix}
     0 & 0 & 0 & 0 & 0 & 0 & 0 & 0 & -3/136 \\
     0 & -1/34 & 0 & -1/34 & 0 & 0 & 0 & 0 & 0 & \\
     1/34 & 0 & 0 & 0 & -1/34 & 0 & 0 & 0 & 0
\end{smallmatrix}\right), \\
\\
&X'^{123}_{(3)}=\left(\begin{smallmatrix}
     0 & 0 & 0 & 0 & -1/34 & 0 & 1/34 & 0 & 0 \\
     0 & 0 & 0 & -1/34 & 0 & 0 & 0 & -1/34 & 0 & \\
     0 & 0 & -3/136 & 0 & 0 & 0 & 0 & 0 & 0
\end{smallmatrix}\right).
\end{array}
\end{equation}

By Eq.(\ref{xa}) and Eq.(\ref{xa'}), we see that $\mathcal{X}'_{1}=\mathrm{I}\mathcal{X}_{1}$ and $\mathcal{X}'_{3}=\mathrm{I}\mathcal{X}_{3}$. With
the factor matrices of Eq.(\ref{x123}) and Eq.(\ref{x'123}), we then see that
$O_1=\mathrm{I},\ O_2=\left(\begin{smallmatrix}
     0 & 0 & -1\\
     0 & 1 & 0\\
     1 & 0 & 0
\end{smallmatrix}\right),$ and $O_3=\mathrm{I}$. Then
$X'_{(1)}=O_1X_{(1)}(O_3\otimes O_2)^{\mathrm{t}}$, $X'_{(2)}=O_2X_{(2)}(O_3\otimes O_1)^{\mathrm{t}}$, and $X'_{(3)}=O_3X_{(3)}(O_2\otimes O_1)^{\mathrm{t}}$, i.e., $(O_1\otimes O_2\otimes O_3)\mathcal{X}_{123}=\mathcal{X}'_{123}$. In fact,
 $\rho$ and $\tau$ are local unitary equivalent:
 $\tau=(U_1\otimes U_2\otimes U_3)\rho(U_1\otimes U_2\otimes U_3)^{\dagger}$, where $U_1=U_3=\mathrm{I}_{2\times 2}$ and $U_2=\frac{1}{2}\left(\begin{smallmatrix}
     1+i & 1+i \\
     -i-1 & 1+i
\end{smallmatrix}\right)$.

\section{Low-rank approximation and LU equivalence}
It is an NP-hard problem to determine the rank of a given tensor. Often, one uses a procedure to calculate CP with a given number of components, such as $R=1,2,3,\cdots$, until the result is good enough. This is a challenge and there are not always possible to find a perfect procedure for fitting CP with given components since the
data of the tensor are not noise-free. Bro and Kiers \cite{BK} have introduced an efficient procedure called CORCONDIA to compare
different numbers of components.

To get a CP decomposition of R components, one often uses the alternating least squares (ALS) given by Carroll, Chang \cite{ccp} and Harshman \cite{Har}.

In the following we use the third-order tensors to show the procedure. Let $\mathcal{X}\in\mathbb{R}^{\alpha\times\beta\times\gamma}$,
we compute an $R$-component nonnegative CANDECOMP/PARAFAC factor model
\begin{equation}\label{approx}
\mathcal{X}\sim\mathcal{W}=\sum^{R}_{r=1}\lambda_{r}\mathbf{a}_{r}\circ\mathbf{b}_{r}\circ\mathbf{c}_{r},
\end{equation}
where $\mathbf{\lambda}\in\mathbb{R}^{R}$ is the weight vector normalizing the columns of $A$, $B$ and $C$ as length one. The error of $\mathcal{R}$ is measured by the loss function or the norm $\|\mathcal{X}-\mathcal{W}\|$. The best low-rank tensor approximation $\mathcal{W}$ is given by
\begin{equation}\label{lf}
\min F(\mathcal{W};\mathcal{X})=\min\limits_{\mathcal{W}} \|\mathcal{X}-\mathcal{W}\|
\equiv\sum^{\alpha}_{i_1=1}\sum^{\beta}_{i_2=1}\sum^{\gamma}_{i_3=1}(x_{i_1i_2i_3}-w_{i_1i_2i_3})^{2},
\end{equation}

The main idea of ALS is to 
fix all but one matrix and then perform an alternating loop until some convergence criterion is satisfied \cite{ccp, Har}. The execution of one loop
goes as follows:
initialize $A$, $B$, and $ C$ and fixing $B$ and $C$. The error function is then $\min\limits_{\tilde{A}}\|X_{(1)}-\tilde{A}(C\odot B)^{\mathrm{t}}\|_{\mathrm{F}}$, where $\tilde{A}=A\cdot \mathrm{diag}(\mathbf{\lambda})$. Using matrix products, one gets an
optimal solution $\tilde{A}=X_{(1)}[(C\odot B)^{\mathrm{t}}]^{\dagger}=X_{(1)}(C\odot B)(C^{\mathrm{t}}C\ast B^{\mathrm{t}}B)^{\dagger}$, then
normalize the columns of $\tilde{A}$ to get $A$. Similarly fixing $A$ and $C$ to solve for $B$, then fixing $A$ and $B$ to solve for $C$. Repeat the procedure until the fitting ceases to improve or maximum iterations exhausted.

The following theorem is based on the conventional ALS method.

\begin{theorem}\label{cpapprox} Suppose two generic 3-qubit states $\rho$ and $\rho'$ are local unitary equivalent. Let $\mathcal{X}$ and $\mathcal{X}'$ be the
CP factorization with given rank $R$ respectively.
Then, the minimum values of their loss functions are the same, i.e.,
\begin{eqnarray}\label{Thap}
\min F(\mathcal{W};\mathcal{X})=\min F(\mathcal{W}';\mathcal{X}').
\end{eqnarray}
where $\mathcal{W}$ is the low rank tensor.
\end{theorem}
\noindent
{\bf{Proof:}} Suppose $(U_1\otimes U_2\otimes U_3)\rho(U_1\otimes U_2\otimes U_3)^t=\rho'$ for $U_i\in \mathrm{SU}(2)$.
Then the corresponding coefficient tensors obey $(O_1\otimes O_2\otimes O_3)\mathcal X=\mathcal X'$ for some
$O_i\in \mathrm{SO}(4)$. Note that $\| \mathcal X\|=\| X_{(1)}\|_F$ for the matricization,
and $X_{(1)}'=O_1(X_{(1)})(O_3\otimes O_2)^t$. Therefore
\begin{align*}
\| \mathcal W-\mathcal X\|&=\|W_{(1)}-X_{(1)}\|_F=\|O_1(W_{(1)})(O_3\otimes O_2)^t-X_{(1)}'\|\\
&=\| \mathcal W'-\mathcal X'\|
\end{align*}
where $\mathcal W'=(O_1\otimes O_2\otimes O_3)\mathcal W$.
Taking minimum over all lower-rank tensor $\mathcal W$, we have shown the result.
$\hfill\Box$

Next we improve the approximation of CP decomposition with orthogonal rank one tensors.
Let $\mathcal{W}=\sum^{R}_{r=1}\lambda_{r}{\mathbf{a}^{(1)}_{r}\circ\mathbf{a}^{(2)}_{r}\circ\mathbf{a}^{(3)}_{r}}$ be a CP decomposition with factor matrices $A_1, A_2, A_3$.
By \cite{olo} there exists a CP decomposition such that
the summands $H_r=\mathbf{a}^{(1)}_{r}\circ\mathbf{a}^{(2)}_{r}\circ\mathbf{a}^{(3)}_{r}$ are orthogonal in the sense that
\begin{equation}\label{muorth}
\langle H_{i}, H_{j}\rangle=\prod^{3}_{l=1}\langle \mathbf{a}^{(l)}_{i},\mathbf{a}^{(l)}_{j}\rangle=\delta_{ij}.
\end{equation}
Ref.{\cite{olo}} provides a modified ALS algorithm with the regression step for orthogonal factor matrices and the polar decomposition.
This improvement allows the orthogonal ALS method converges globally.
The following result is clear.

\begin{theorem}\label{cpappr}  Let $\rho$ and $\rho'$ be two locally unitary equivalent 3-qubit states with orthogonal rank $R$ CP decompositions $\mathcal W$ and $\mathcal W'$ with the coefficient tensors $\mathcal X$ and $\mathcal X'$ respectively.  Then they have the same minimum values
of the cost function and there exist $O_1,O_2,O_3\in SO(4)$, such that $(O_1\otimes O_2\otimes O_3)\mathcal{W}=\mathcal{W}'$.
\end{theorem}

The following result offers a sufficient condition.
\begin{corollary}\label{OCp} Let $\rho, \rho'$ be two generic 3-qubit states with
the factor matrices $A, B, C$ and $A', B', C'$ in orthogonal rank-$R_{123}$ CP approximations respectively. Suppose there are
$O_1, O_2, O_3\in SO(3)$ such that
$A'^{123}=O_1A^{123},\ B'^{123}=O_2B^{123}$ and $C'^{123}=O_3C^{123}$ as well as $\mathcal{X}'_{j_1\cdots j_M}=(O_{j_1}\otimes\cdots\otimes O_{j_M})\mathcal{X}_{j_1\cdots j_M}$.
Then $\rho \overset{LOU}{\simeq}\rho'$.
\end{corollary}

\begin{corollary}\label{OCpp}
Let $\rho, \rho'$ be two generic 3-qubit states with orthogonal CP approximations
with factor matrices $\mathcal W_{i_1\cdots\i_r}$ and $\mathcal W_{i_1\cdots\i_r}'$ associated with their coefficient tensors $\mathcal X$ and $\mathcal X'$ respectively. If
$(A^{j_1\cdots j_M}_{j_i})^{\mathrm{t}}A^{j_1\cdots j_M}_{j_i}=(A'^{j_1\cdots j_M}_{j_i})^{\mathrm{t}}{A}'^{j_1\cdots j_M}_{j_i}$ and $(A^{123}_{j_i})^{\mathrm{t}}{A^{j_1\cdots j_M}_{j_i}}=(A'^{123}_{j_i})^{\mathrm{t}}{A}'^{j_1\cdots j_M}_{j_i}(M\leq 3,1\leq j_1<\cdots<j_M\leq 3,i=1,\cdots,M)$,
then $\rho$ and $\rho'$ are local unitary equivalent.
\end{corollary}
{\bf{Proof:}} The result follows from
Theorem {\ref{h}} and Theorem {\ref{cpappr}}.
$\hfill\Box$

\section{Conclusion}

We have studied the local unitary equivalence for generic multi-qubit states using a new method of the CP decomposition in higher-order tensor theory. The CP decomposition is one of the most important generalizations of the spectral decomposition in the matrix theory. The quantum states can be cast in the Bloch representation, so bring in critical mathematical study of tensors is natural. In this paper, we have formulated the CP decomposition method in the context of entanglement and local unitary equivalence, which are incredibly notions to study in quantum computation and quantum information. Explicitly we have shown that several key invariants of the CP decompositions are also quantum invariants for local unitary equivalence, in particular, the rank, factor matrices, orthogonal face matrices etc. are confirmed. We then found some criteria to judge
local unitary equivalence between two quantum states, both for tripartite and multi-partite states in general. We hope our study can lead to further discussions in deciphering the mystery in local unitary equivalence or
other important questions in quantum computation.


\addcontentsline{toc}{chapter}{Acknowledgment}
\section*{Acknowledgments}
The research is supported in part by Simons Foundation grant no. 523868 and National Natural Science Foundation grant nos. 12126351 and 12126314.

\end{document}